\documentclass[a4paper]{article}

\usepackage{INTERSPEECH2022}
\title{3M: Multi-loss, Multi-path and Multi-level Neural Networks for speech recognition}
\name{Zhao You$^{*1}$, Shulin Feng$^{*1}$, Dan Su$^1$, Dong Yu$^2$}
\address{$^1$Tencent AI Lab, Shenzhen, China\\
$^2$Tencent AI Lab, Bellevue, WA, USA }
\email{\{dennisyou, shulinfeng, dansu, dyu\}@tencent.com}

\begin{document}

\maketitle

\begin{abstract}
Recently, Conformer based CTC/AED model has become a mainstream architecture for ASR. In this paper, based on our prior work, we identify and integrate several approaches to achieve further improvements for ASR tasks, which we denote as multi-loss, multi-path and multi-level, summarized as "3M" model. Specifically, multi-loss refers to the joint CTC/AED loss and multi-path denotes the Mixture-of-Experts(MoE) architecture which can effectively increase the model capacity without remarkably increasing computation cost. Multi-level means that we introduce auxiliary loss at multiple level of a deep model to help training.
We evaluate our proposed method on the public WenetSpeech dataset and experimental results show that the proposed method provides $ 12.2\% \sim 17.6\% $ relative CER improvement over the baseline model trained by Wenet toolkit. On our large scale dataset of 150k hours corpus, the 3M model has also shown obvious superiority over the baseline Conformer model. Code is publicly available at https://github.com/tencent-ailab/3m-asr.

\end{abstract}

\renewcommand{\thefootnote}{\fnsymbol{footnote}}
\footnotetext[1]{Equal contribution.}

\section{Introduction}
End-to-End(E2E) automatic speech recognition(ASR) have gained large improvements in recent years. As a combination of convolution module and self-attention mechanism, Conformer\cite{conformer} can model local and global dependencies of an audio sequence and achieve a better performance, making it become a favored choice to train benchmarks for many E2E ASR toolkits(e.g. Espnet\cite{watanabe2018espnet}, Wenet\cite{zhang2021wenet}). CTC/attention-based encoder-decoder, also known as CTC/AED, is a popular E2E ASR framework, which effectively utilizes advantages of both architectures to improve robustness and achieve faster convergence. Previous works\cite{DBLP:journals/corr/HoriWZC17,wu2019improving,watanabe2017hybrid,hori2017joint} have proved this multi-loss training can significantly improve the performance of E2E ASR systems.


Scaling up to a larger model has been an effective way towards a flexible and powerful E2E ASR system. It has shown that a large Conformer model(e.g. SpeechStew\cite{chan2021speechstew}) can achieve state-of-the-art(SoTA) results across a wide variety of tasks. However, developing large models in real-world application is seriously hindered by the expensive computation cost of both training and inference time. Several techniques have been proposed to reduce the model complexity, such as knowledge distillation, low-rank decomposition, quantization and pruning, but these methods inevitably suffer from performance degradation.


Recently, mixture of experts (MoE) based approaches \cite{jacobs1991adaptive, jordan1994hierarchical} have been intensively investigated and applied in different tasks such as language modeling \cite{lepikhin2020gshard, fedus2021switch}, image classification \cite{gross2017hard, ahmed2016network, wang2020deep, cai2021dynamic}, and speech recognition \cite{you2021speechmoe,you2021speechmoe2}.
MoE models can easily increase the model capacity by increasing the number of experts. With the introduction of the sparsely-gated mixture-of-experts layer\cite{shazeer2017outrageously}, MoE models can dynamically route inputs to corresponding expert networks, which enables us to satisfy training and inference efficiency by having sub-network activated on per-example basis.

Based on the joint CTC/AED multi-loss training framework, we explore the MoE approach on the Conformer model, named Conformer-MoE, where the second Macaron-style half-step feed-forward layer of each Conformer block is modified into the MoE layer. With sparsely-gated mechanism, the MoE layer will route their inputs to the top-1 expert with largest route probability, thus keeping the computational cost roughly constant. The Conformer-MoE activate different sub-networks when dealing with different input samples and we call this property as multi-path. Additionally, we further propose to integrate separate attention decoders at multiple intermediate level of encoder layers, which effectively accelerate convergence and get a better performance. The multi-level property, along with multi-path and multi-loss, makes up our final 3M model


The rest of the paper is organized as follows. Section 2 reviews the previous work of Conformer and SpeechMoE, and Section 3 presents our proposed method 3M. The experimental results are reported in Section 4. Finally, we conclude this paper in Section 5.

\label{sec:typestyle}
\begin{figure*}[!tb]
\begin{minipage}[b]{0.9\linewidth}
  \centering
  \centerline{\includegraphics[width=14.5cm,height=11cm]{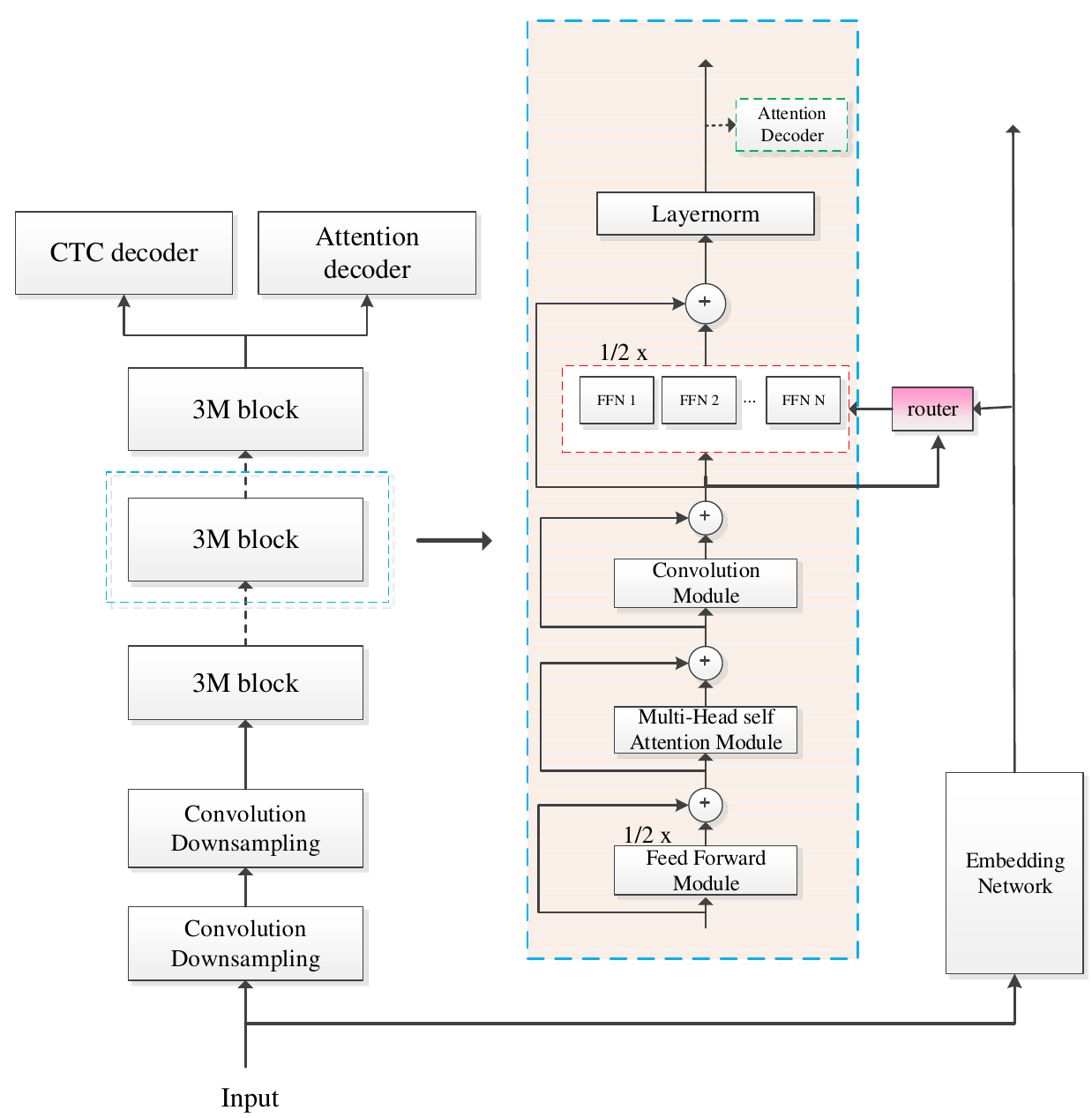}}
\end{minipage}
\caption{
illustration for the architecture of 3M model, which is composed of a shared encoder with several 3M blocks, a CTC decoder, an Attention Decoder and an embedding network.
}
\end{figure*}

\section{Previous works}

This section mainly describes previous work of Conformer and SpeechMoE.
\subsection{Conformer}
The Conformer is proposed by \cite{conformer}, and each Conformer block comprises two Feed Forward modules sandwiching the Multi-Head self-attention module and the Convolution module. Given the input $x$, we can get the output $y$ of Conformer block:
\begin{equation}
\hat{x}= x + \frac{1}{2}FFN(x)
\end{equation}
\begin{equation}
\widetilde{x}= \hat{x} + MHSAN(\hat{x})
\end{equation}
\begin{equation}
\overline{x}= \widetilde{x} + Conv(\widetilde{x})
\end{equation}
\begin{equation}
y= LN(\overline{x} + \frac{1}{2}FFN(\overline{x}))
\end{equation}
where FFN represents the feed-forward module, MHSAN represents the multi-head self-attention module, Conv represents the convolution module, and LN represents the layer norm.
\subsection{SpeechMoE}
The SpeechMoE is proposed in \cite{you2021speechmoe}, which comprises multiple MoE layers, non-expert layers and a shared embedding network. Each MoE layer consists of $n$ experts and a router. It takes the output of the previous layer
and the shared embedding as input and routes each speech frame to the top-1 expert with the largest route probability. Let $W^{l}_{r}$, $e^{c}$ and $o^{l-1}$ be the router weights of the $l$-th layer, shared embedding and the output of the previous layer, then the router probability can be defined as follows:

\begin{equation}
  r^{l} = W_{r}^{l} \cdot Concat(e^{c};o^{l-1})
\end{equation}

\begin{equation}
  {p^{l}_{i}} = \frac{exp^{r^{l}_{i}}}{\sum_{j=1}^{n}exp^{r^{l}_{j}}}
\end{equation}
Then, the selected expert's output is also gated by router probability to get the output of the MoE layer,
\begin{equation}
y^{l} = p^{l}_{i} E^{l}_{i}
\end{equation}

Since only one expert is active in each layer, the SpeechMoE can keep the computational cost constant while scaling up to a very large model.
To achieve better sparsity and balance among different experts, the sparsity L1 loss $L_{s}$ and mean importance loss $L_{m}$ are added into the loss function:

\begin{equation}
L_{s}= \frac{1}{k}\sum_{j=1}^{k} \parallel{\hat{p}_{j}}\parallel_{1}
\end{equation}

\begin{equation}
{L}_{m}= n\sum_{i=1}^{n} ({\frac{1}{k}\sum_{j=1}^{k} p_{ij}})^{2}
\end{equation}
where $\hat{p}_{j}$ stands for the unit-normalized router probability distribution of frame $j$, and $k$ is the number of frames in this mini-batch. $p_{ij}$ stands for the router probability on expert $i$ of frame $j$.

\newcommand{\tabincell}[2]{\begin{tabular}{@{}#1@{}}#2\end{tabular}}

\begin{table*}
\caption{\textit{Results of Conformer, Conformer-MLevel, Conformer-MoE and 3M. } }
\label{tab:1}
\begin{center}
\scalebox{1.1}{
\begin{tabular}{ccccccc}
\toprule[2pt]
{Testset} & Confomer & Conformer-MLevel& Conformer-MoE  (16e) & 3M  (16e) & 3M (32e) & 3M (64e)\\
 \hline
 cctv & 1.60 & 1.52 &1.48 & 1.47 & 1.43& 1.42\\
  deyunshe & 16.48 & 16.01 & 15.56 & 15.20 & 14.78 & 14.31\\
  liyongle & 4.63 & 3.96 &3.71 & 3.53 & 3.50 & 3.37\\
  luoxiang & 3.93 & 3.51 &3.30 & 3.11 & 3.10 & 3.08\\
  luozhengyu & 2.78 & 2.55 & 2.4 & 2.30 & 2.27 & 2.19\\
  qichezhijia & 6.89 & 6.58 & 6.42 & 6.34 & 6.30 & 6.25\\
  shenghuo & 9.84 & 9.49 & 9.16 & 9.03 & 8.41 & 8.13\\
  tianxiazuqiu & 3.82 & 3.77 & 3.59 & 3.52 & 3.46 & 3.37\\
  wangzhe & 11.75 &11.29& 10.87 & 10.63 & 10.43 & 10.36\\
  xiaoaidashu & 5.94 & 5.83 & 5.81 & 5.74 & 5.70 & 5.65\\
\bottomrule[2pt]
\end{tabular}
}
\end{center}
\end{table*}
\vspace{10pt}
\section{3M model}
\subsection{Multi-loss}
To date, CTC/AED is commonly used in E2E ASR, which effectively utilizes the
advantages of both architectures in training and decoding. On one hand, the attention lacks left-to-right constraints, making it difficult to generate proper alignments in the case of noisy data and/or long input sequences. Augmenting the constrained CTC loss based on the forward-backward algorithm can greatly reduce the number of irreuglarly aligned utterances. On the other hand, CTC imposes the conditional independence constraint that output predictions are independent, which is not true for ASR. Attention objective can help eliminate the conditional independence constraint and back propagate the context information to rectify the probability distribution output by CTC.
When employing decoding, both attention-based scores and CTC scores can be combined in a rescoring/one-pass beam search to improve the performance. We regard the CTC/AED joint training as multi-loss property. As shown in Figure 1, the proposed model has a CTC branch and an attention branch with a shared encoder.

\subsection{Multi-path}
Here, we propose to apply MoE architecture on Conformer model. Specifically, as shown in Figure 1, we extend the second Macaron-style feed-forward module to the MoE layer, which consists of $N$ FFN module and a router. Similar to SpeechMoE\cite{you2021speechmoe}, each FFN module stands for one expert network and the router takes the shared embedding and the output from the previous module as input and routes each speech frame to the top-1 expert with the largest route probability. Expert outputs are also gated by the corresponding router probability. With this sparse-gated mechanism, the proposed model can activate different sub-networks for different samples, which is regarded as multi-path property. We have also attempted to modify both FFN modules in the Conformer block into MoE layers, but it fails to achieve a better performance.


\subsection{Multi-level}
As we know, the lower layers of the Conformer encoder also have a low  correlation with grapheme information. Generally, the input of the attention decoder is the top layer of the Conformer encoder. In order to make the lower layers of encoder have a earlier access to grapheme information from the decoder, we propose to integrate separate attention decoders at intermediate layers of the encoder, as shown in green box part of Figure 1. In our practices, two additional decoders are applied to intermediate layers at 1/3 depth and 2/3 depth of the encoder, which are only used for training with AED loss and ignored during inference.


\subsection{ Training objective}
\subsubsection{ MoE training loss}
We have introduced the mentioned $L_{s}$ and $L_{m}$ loss to make the router probabilities sparse and diverse. To provide
enough distinction information for routers, the shared embedding network which converts low-level features to high-level
embedding is necessary to help the router attain better selecting effect.  The embedding training loss provides reliable embeddings for the routers. The training loss of MoE is defined as:
\begin{equation}
L_{MoE}=\alpha L_{s}(x) + \beta {L}_{m}(x) + \gamma  L_{e}(x;y)
\end{equation}
Where $L_{s}$ and ${L}_{m}$ are the mentioned sparsity $L1$ loss and mean importance loss, used to encourage sparsity and diversity of the model. The embedding loss $L_{e}$ is the CTC loss. $\alpha$, $\beta$, and $\gamma$ is the scale for $L_{s}$, ${L}_{m}$ and $L_{e}$ respectively.
\subsubsection{ Joint CTC/AED training loss}
In this paper, the CTC loss and multi-level AED loss are combined in the training of the 3M model:
\begin{small}
\begin{equation}
\begin{split}
L_{Joint}=\eta L_{c}(x;y)+(1-\eta)\sum_{j=1}^{K} L_{a_{j}}(x;y)
\end{split}
\end{equation}
\end{small}
Among these items, $L_{c}$ is the CTC loss. $K$ is the number of level. $L_{a_{j}}$ is the AED loss for the  transformer decoder of $j$-th level.  $\eta$ are the interpolated weight for CTC loss and AED loss.
\subsubsection{Loss function}
Given the input $x$, grapheme target $y$, the complete loss function of our method is defined as:
\begin{equation}
L(x;y)= L_{MoE}+L_{Joint}
\end{equation}

\vspace{10pt}
\section{EXPERIMENTS}
\subsection{WenetSpeech Task}
\label{sec:format}
We first evaluate our proposed method on the WenetSpeech\cite{zhang2022wenetspeech} Dataset, which a is multi-domain Mandarin corpus consisting of 10005 hours of high-quality labeled speech.
Evaluation is performed in terms of word error rate (WER) on the Dev, Test\underline{~}Net and Test\underline{~}Meeting, described in \cite{zhang2022wenetspeech}.
The input speech uses 80-dimension log-Mel filterbank features, which are computed with a 25ms window and shifted every 10ms.  Spec-Augment is applied 2 frequency masks with maximum frequency mask (F = 30) and 2 time masks with maximum time mask (T = 50) to alleviate over-fitting. A global mean and variance normalization is used as data preparation. The max number of epochs is 26. We set the checkpoint with the lowest CTC loss as the final model.

We use the benchmarks from Kaldi, Espnet and Wenet listed in \cite{zhang2022wenetspeech} as our baselines. Our Conformer model uses the same setup as Espnet and Wenet benchmarks, which consists of 12-block Conformer encoder($d^{ff}=2048, H=8, d^{att}=512, CNN_{kernel}=15$) and 6-block transformer decoder($ d^{ff}=2048, H=8 $). As for the MoE layer, we set the number of experts to be 16, 32 and 64, noted as 16e, 32e and 64e respectively. The shared embedding network is a static model without MoE layer but a similar structure to the baseline, which contains 6 Conformer encoder layers and is pretrained with CTC objective before used for MoE training. The whole model is trained with CTC and attention objectives, along with our mentioned auxiliary losses. For all experiments on MoE models, we set the hyper-parameters $\alpha=0.15, \beta=0.15, \gamma=0.01, \eta=0.3$. A set of  5535 Mandarin characters and 26 English letters is used as the modeling units. For decoding, we generate the N-Best hypotheses by the CTC decoder and rescore them by the attention decoder to get the final results.

As shown in Table 2, our Conformer-MoE models, equipped with the multi-path property, achieve consistent performance improvements on three test sets over Kaldi, Espnet and Wenet benchmarks. Moreover, by increasing the number of experts, Conformer-MoE gets a better performance on Dev but 64e doesn't get consistent improvements on Test\_Net and Test\_Meeting, which have more difficult domain and unmatched data. In total, Conformer-MoE(32e) achieves the best performance on these test sets. Compared with Wenet benchmark, it provides $12.2\% \sim 17.6\%$ relative character error rate(CER) reduction. Our 3M model, Comformer-MoE with additional multi-level property, doesn't get a better performance over Conformer-MoE so we did not post its results. 

\subsection{150,000 hours Task}
In this experiment, we scale up the size of training corpus to 150000 hours to further identify the effectiveness of our proposed method. The training corpus is collected from different application domains. In order to improve the system robustness, we simulate six types of different mixing conditions like voice processing(vp), acoustic echo cancellation(aec), reverberation(rir) and signal-to-noise-ratio(SNR), etc. The SNR is set between 15 and 30 dB, and the reverberation time is set between 0 and 900 milliseconds. The test sets are collected from YouTube, consisting of 15 hours of audio. They add up to 65536 utterances and are divided into 10 domains, covering game commentary, documentary, comic dialogue and so on.

Our baseline Conformer model is implemented by Wenet toolkits and it consists of two convolution downsampling layers, a 18-block Conformer encoder and a 4-block bi-transformer decoder (2-block for forward and 2-block for backward). For MoE models, the number of experts are also set to be 16, 32 and 64. The shared embedding network is similar as before and has 7 Conformer blocks. For the multi-level setting, two separate bi-transformer decoders are applied to the intermediate layers at 1/3 depth and 2/3 depth of the encoder. Since decoders at intermediate level are only used for training, there is no increase computational cost in decoding for 3M models. All model are trained with CTC and attention objectives and hyper-parameters settings are the same as before. We use the floating point operations(FLOPs) for a one-second example to evaluate the inference computation cost and Table 3 shows the comparison on parameters amount and FLOPs for Conformer, Conformer-MoE and 3M.

Experimental results are shown in Table 1.
It is clear to see from column 2 and column 5 that 3M (16e) achieves lower character error rate than the baseline Conformer model and the gain is 11.5\% on average.
With the increase of data amount, MoE models with more experts are easier to get a better performance. We can see that 3M model achieves consistent improvements from 16e to 64e on all test sets in 150000 hours task, which is different from the WenetSpeech results. In total, 3M(64e) provides 14.9\% relative CER improvement on average over the baseline Conformer model.

To analyze the efficacy of 3M model, we must study how much the multi-path and multi-level contribute to the performance individually.  We train Conformer-MoE (multi-path) and Conformer-MLevel (multi-level) model by only keeping the multi-path and the multi-level in 3M model, respectively. In Table 1, we see that a significant gain benefits from both the multi-level and multi-path. Especially, compared with the baseline model, the gain is 9.2\% (from column 2 and column 4) for the multi-path and 5.6\% (from column 2 and column 3) for the multi-level. It suggests that  both the multi-level and multi-path are critical to the quality of the proposed 3M model.

\begin{table}
\caption{\textit{CER results on WenetSpeech for public benchmarks and Conformer-MoE} }
\label{tab:1}
\begin{center}
\scalebox{1.0}{
\begin{tabular}{cccc}
\toprule[2pt]
{Toolkit} & Dev & Test\underline{~}Net & Test\underline{~}Meeting \\
 \hline
 Kaldi & 9.07 & 12.83 & 24.72\\
  Espnet & 9.70 & 8.90 & 15.90\\
   Wenet & 8.88 & 9.70 & 15.59\\
   Conformer-MoE (16e) & 7.67 & 8.28 & 13.96\\
    Conformer-MoE (32e) & 7.49 & \textbf{7.99} & \textbf{13.69}\\
    Conformer-MoE (64e) & \textbf{7.19} & 8.36 & 13.72\\
\bottomrule[2pt]
\end{tabular}
}
\end{center}
\end{table}

\begin{table}
\caption{\textit{A comparsion of Conformer, Conformer-MoE and 3M. } }
\label{tab:1}
\begin{center}
\begin{tabular}{ccc}
\toprule[2pt]
{Model} & Params & FLOPs \\
 \hline
 Conformer & 120M & 8.3B \\
  Conformer-MoE (16e) & 425M & 12.3B \\
   3M (16e) & 500M & 12.3B \\
   3M (32e) & 775M & 12.3B \\
   3M (64e) & 1.37B & 12.3B \\
\bottomrule[2pt]
\end{tabular}
\end{center}
\end{table}

\section{Conclusions and future work}
In this paper, we introduce the 3M model for speech recognition. Specifically, based on the CTC/AED E2E ASR framework, we apply MoE on the Conformer model and apply AED at multiple levels of the model. We demonstrated the effectiveness of our proposed method on both the public WenetSpeech dataset and a large scale dataset.
Experimental results show that the Conformer-MoE provides up to 17.6\% relative CER improvement compared with the Wenet benchmark on the WenetSpeech dataset.
The mutli-level property can also contribute to a better performance. Overall, our 3M model achieves 14.9\% relative CER improvement on average over the baseline Conformer model on the large scale dataset.  Future work
includes increasing the number of experts by one or two orders of magnitudes, and exploring the proposed 3M model with other end-to-end
training framework such as transformer transducers.


\bibliographystyle{IEEEtran}

\bibliography{mybib}

\end{document}